\documentstyle[preprint,aps]{revtex}

\begin{document}

\draft
\title
{Fractal index, central charge and fractons}

\author
{Wellington da Cruz\footnote{E-mail: wdacruz@exatas.uel.br} 
and Rosevaldo de Oliveira}

\address
{Departamento de F\'{\i}sica,\\
 Universidade Estadual de Londrina, Caixa Postal 6001,\\
Cep 86051-970 Londrina, PR, Brazil\\}
 
\date{\today}

\maketitle

\begin{abstract}

We introduce the notion of fractal index associated with the universal 
class $h$ of particles or quasiparticles, termed fractons, which obey 
specific fractal statistics. A connection between fractons and 
conformal field theory(CFT)-quasiparticles is established taking into 
account the central charge $c[\nu]$ and the particle-hole duality 
$\nu\longleftrightarrow\frac{1}{\nu}$, for integer-value $\nu$ of the 
statistical parameter. In this way, we derive the Fermi velocity in 
terms of the central charge as $v\sim\frac{c[\nu]}{\nu+1}$. The 
Hausdorff dimension $h$ which labelled the universal classes of 
particles and the conformal anomaly are therefore related. 
Following another route, we also established a connection between Rogers 
dilogarithm function, Farey series of rational numbers and the 
Hausdorff dimension.

\end{abstract}


We consider the conformal field theory(CFT)-quasiparticles 
( edge excitations ) in connection with the concept 
of fractons introduced in\cite{R1}. These excitations have been considered at 
the edge of the quantum Hall systems which in the fractional regime assume the 
form of a chiral Luttinger liquid\cite{R2}. Beyond this, 
conformal field theories have been exploited in a variety of contexts, 
including statistical mechanics at the critical 
point, field theories, string theory, and in various 
branches of mathematics\cite{R3}.

In this Letter, we suppose that the fractal statistics obeyed by 
fractons are shared by CFT-quasiparticles. Thus, the central charge, a 
model dependent constant is related to the universal class $h$ of the 
fractons. We define the {\it fractal index} associated with 
these classes as 

\begin{equation}
\label{e.1}
i_{f}[h]=\frac{6}{\pi^2}\int_{\infty(T=0)}^{1(T=\infty)}
\frac{d\xi}{\xi}\ln\left\{\Theta[\cal{Y}(\xi)]\right\}
\end{equation}

\noindent

and after the change of variable $\xi=x^{-1}$, we obtain

\begin{equation}
\label{e.2}
i_{f}[h]=\frac{6}{\pi^2}\int_{0}^{1}
\frac{dx}{x}\ln\left\{\Theta[{\cal{Y}}(x^{-1})]\right\}
\end{equation}

\noindent where in the Eq.$(\ref{e.1})$

\begin{eqnarray}
\Theta[{\cal{Y}}]=
\frac{{\cal{Y}}[\xi]-2}{{\cal{Y}}[\xi]-1}
\end{eqnarray}

\noindent is the single-particle partition function of the 
universal class $h$ and $\xi=\exp\left\{(\epsilon-\mu)/KT\right\}$, 
has the usual definition. The function ${\cal{Y}}[\xi]$ satisfies 
the equation 

\begin{eqnarray}
\label{e.46} 
\xi=\left\{{\cal{Y}}[\xi]-1\right\}^{h-1}
\left\{{\cal{Y}}[\xi]-2\right\}^{2-h}.
\end{eqnarray}

We note here that the general solution of the algebraic 
equation derived from this last one is of the form

\[ 
{\cal{Y}}_{h}[\xi]=f[\xi]+{\tilde{h}}
\]

or

\[ 
{\cal{Y}}_{\tilde{h}}[\xi]=g[\xi]+h,
\]

\noindent where ${\tilde{h}}=3-h$, is a 
{\it duality symmetry}\footnote{This means that fermions($h=1$) 
and bosons($h=2$) are dual objects. As a result we have a {\it 
fractal supersymmetry}, since for the particle with spin $s$ 
within the class $h$, its dual $s+\frac{1}{2}$ is within the class ${\tilde{h}}$.} 
between 
the classes. The functions $f[\xi]$  and $g[\xi]$ at least 
for third, fourth degrees algebraic equation differ by plus and minus 
signs
in some terms of their expressions.

The particles within each class $h$ satisfy 
specific {\bf fractal statistics}\footnote{In fact, we have here 
{\it fractal functions} as discussed in\cite{R4}. }\cite{R1}

\begin{eqnarray}
\label{e.h} 
n&=&\xi\frac{\partial}{\partial{\xi}}\ln\Theta[{\cal{Y}}]\nonumber\\
&=&\frac{1}{{\cal{Y}}[\xi]-h}
\end{eqnarray}

\noindent and the fractal parameter\footnote{This parameter 
describes the properties  of the path
({\it fractal curve}) of the quantum-mechanical particle.}  
(or Hausdorff dimension) 
$h$
defined in the interval $1$$\;$$ < $$\;$$h$$\;$$ <$$\;$$ 2$ 
is related to the  spin-statistics relation $\nu=2s$ 
through the {\it fractal spectrum}

\begin{eqnarray}
\label{e.7}
&&h-1=1-\nu,\;\;\;\; 0 < \nu < 1;\;\;\;\;\;\;\;\;
 h-1=\nu-1,\;
\;\;\;\;\;\; 1 <\nu < 2;\\
&&etc.\nonumber
\end{eqnarray}

\noindent For $h=1$ we have fermions, with ${\cal{Y}}[\xi]=\xi+2$, 
$\Theta[1]=\frac{\xi}{\xi+1}$ and $i_{f}[1]=
\frac{6}{\pi^2}\int_{\infty}^{1}\frac{d\xi}{\xi}
\ln\left\{\frac{\xi}{\xi+1}\right\}=\frac{1}{2}$. 
For $h=2$ we have bosons, with  
${\cal{Y}}[\xi]=\xi+1$, $\Theta[2]=
\frac{\xi-1}{\xi}$ and $i_{f}[2]=\frac{6}{\pi^2}\int_{\infty}^{1}\frac{d\xi}{\xi}
\ln\left\{\frac{\xi-1}{\xi}\right\}=1$. On the other hand, 
for the universal class 
$h=\frac{3}{2}$, 
we have fractons with ${\cal{Y}}[\xi]=\frac{3}{2}+\sqrt{\frac{1}{4}+\xi^2}$, 
$\Theta\left[\frac{3}{2}\right]=\frac{\sqrt{1+4\xi^2}-1}{\sqrt{1+4\xi^2}+1}$ and  
$i_{f}\left[\frac{3}{2}\right]=\frac{6}{\pi^2}\int_{\infty}^{1}\frac{d\xi}{\xi}
\ln\left\{\frac{\sqrt{1+4\xi^2}-1}{\sqrt{1+4\xi^2}+1}\right\}=\frac{3}{5}$.

The distribution function for each class $h$ above, 
as we can check, are given by

\begin{eqnarray}
n[1]&=&\frac{1}{\xi+1},\\
n[2]&=&\frac{1}{\xi-1},\\
n\left[\frac{3}{2}\right]&=&\frac{1}{\sqrt{\frac{1}{4}+\xi^2}},
\end{eqnarray}

\noindent i.e. we have the Fermi-Dirac distribution, the Bose-Einstein 
distribution and the fracton distribution of the universal 
class $h=\frac{3}{2}$, 
respectively . Thus, our formulation generalizes {\it in a natural 
way} the fermionic and 
bosonic distributions for particles assuming 
rational or irrational values for the spin quantum number $s$. In this way, 
our approach can be understood as a {\it quantum-geometrical} description of 
the statistical laws of Nature. This means that the (Eq.\ref{e.h}) 
captures the observation about the fractal characteristic of the 
{\it quantum-mechanical} path, which reflects the 
Heisenberg uncertainty principle. 

The fractal index as defined has a connection with the central 
charge or conformal anomaly $c[\nu]$, a dimensionless number which 
characterizes conformal field theories in two dimensions. This way, 
we verify that 
the conformal anomaly is associated with universality classes, 
i.e. universal classes $h$ 
of particles. Now, we consider the particle-hole duality 
$\nu\longleftrightarrow\frac{1}{\nu}$ for integer-value 
$\nu$ of the statistical parameter in connection 
with the universal class $h$. For bosons and fermions, we have

\[
\left\{0,2,4,6,\cdots\right\}_
{h=2}
\]

and

\[
\left\{1,3,5,7,\cdots\right\}_
{h=1}
\]

such that, the central charge for $\nu$ {\it even} is defined by

\begin{eqnarray}
\label{e.11}
c[\nu]=i_{f}[h,\nu]-i_{f}\left[h,\frac{1}{\nu}\right]
\end{eqnarray}

\noindent and for $\nu$ {\it odd} is defined by

\begin{eqnarray}
\label{e.12}
c[\nu]=2\times i_{f}[h,\nu]-i_{f}\left[h,\frac{1}{\nu}\right],
\end{eqnarray}

\noindent where $i_{f}[h,\nu]$ means the fractal 
index of the universal class $h$ which contains the 
statistical parameter $\nu=2s$ or the particles 
with distinct spin values which obey specific 
fractal statistics. We assume that the fractal 
index $i_{f}[h,\infty]=0$ ( the class $h$ is undetermined ) 
and we obtain, for example, the results

\begin{eqnarray}
&&c[0]=i_{f}[2,0]-i_{f}[h,\infty]=1;\nonumber\\
&&c[1]=2\times i_{f}[1,1]-i_{f}[1,1]=\frac{1}{2};\nonumber\\
&&c[2]=i_{f}[2,2]-i_{f}\left[\frac{3}{2},\frac{1}{2}\right]=
1-\frac{3}{5}=\frac{2}{5};\\
&&c[3]=2\times i_{f}[1,3]-i_{f}\left[\frac{5}{3},\frac{1}{3}\right]=
1-0.656=0.344;\nonumber\\
&&etc,\nonumber
\end{eqnarray}

\noindent where the fractal index for $h=\frac{5}{3}$ is obtained from 

\begin{eqnarray}
&&i_{f}\left[\frac{5}{3}\right]=\frac{6}{\pi^2}\int_{\infty}^{1}\frac{d\xi}{\xi}
\\
&&\times\ln\left\{\frac{{\sqrt[3]{\frac{1}{27}+\frac{\xi^3}{2}
+\frac{1}{18}\sqrt{12\xi^3+81\xi^6}}}+
\frac{1}{9\sqrt[3]{\frac{1}{27}+\frac{\xi^3}{2}+\frac{1}{18}
\sqrt{12\xi^3+81\xi^6}}}-\frac{2}{3}}
{{\sqrt[3]{\frac{1}{27}+\frac{\xi^3}{2}
+\frac{1}{18}\sqrt{12\xi^3+81\xi^6}}}+
\frac{1}{9\sqrt[3]{\frac{1}{27}+\frac{\xi^3}{2}+\frac{1}{18}
\sqrt{12\xi^3+81\xi^6}}}+\frac{1}{3}}\right\}\nonumber\\
&&=0.656\nonumber
\end{eqnarray}

\noindent and for its dual we have

\begin{eqnarray}
&&i_{f}\left[\frac{4}{3}\right]=\frac{6}{\pi^2}\int_{\infty}^{1}\frac{d\xi}{\xi}
\\
&&\times\ln\left\{\frac{{\sqrt[3]{-\frac{1}{27}+\frac{\xi^3}{2}
+\frac{1}{18}\sqrt{-12\xi^3+81\xi^6}}}+
\frac{1}{9\sqrt[3]{-\frac{1}{27}+\frac{\xi^3}{2}+\frac{1}{18}
\sqrt{-12\xi^3+81\xi^6}}}-\frac{1}{3}}
{{\sqrt[3]{-\frac{1}{27}+\frac{\xi^3}{2}
+\frac{1}{18}\sqrt{-12\xi^3+81\xi^6}}}+
\frac{1}{9\sqrt[3]{-\frac{1}{27}+\frac{\xi^3}{2}+\frac{1}{18}
\sqrt{-12\xi^3+81\xi^6}}}+\frac{2}{3}}\right\}\nonumber\\
&&=0.56.\nonumber
\end{eqnarray}

\noindent From the Table I we can observe the correlation 
between the classes $h$ of particles and their fractal index, so our 
approach manifest {\bf a robust} consistence in accordance with the unitary
$c[\nu]$$\;$$ <$$\;$$ 1$ representations\cite{R3}.

\newpage
\vspace{3mm}
\begin{center}
Table I
\end{center}
\begin{center}
\begin{tabular}{|c|c|c|c|c|c|}
\hline
$h$ & $i_{f}[h]$ & Denomination & $\nu$ & $s$ & $c[\nu]=i_{f}[h,\nu]$\\
\hline
$2$ & $1$ & bosons & $0$ & $0$ & $1$\\
\hline
$\cdots$ & $\cdots$ & fractons & $\cdots$ & $\cdots$ & $\cdots$\\
\hline
$\frac{5}{3}$ & $0.656$ & fractons & $\frac{1}{3}$ & $\frac{1}{6}$ & $0.656$\\
\hline
$\cdots$ & $\cdots$ & fractons & $\cdots$ & $\cdots$ & $\cdots$\\
\hline
$\frac{3}{2}$ & $0.6$ & fractons & $\frac{1}{2}$ & $\frac{1}{4}$ & $0.6$\\
\hline
$\cdots$ & $\cdots$ & fractons & $\cdots$ & $\cdots$ & $\cdots$\\
\hline
$\frac{4}{3}$ & $0.56$& fractons & $\frac{2}{3}$ & $\frac{1}{3}$ & $0.56$\\
\hline
$\cdots$ & $\cdots$ & fractons & $\cdots$ & $\cdots$ & $\cdots$\\
\hline
$1$ & $0.5$ & fermions & $1$ & $\frac{1}{2}$ & $0.5$\\
\hline
\end{tabular}
\end{center}
\vspace{5mm}

\noindent Therefore, since $h$ is defined within the interval 
$ 1$$\;$$ < $$\;$$h$$\;$$ <$$\;$$ 2$, the corresponding fractal index 
is into the interval $0.5$$\;$$ < $$\;$$i_{f}[h]$$\;$$ <$$\;$$ 1$. Howewer, 
the central charge $c[\nu]$ can assumes values less than $0.5$. Thus, we 
distinguish two concepts of central charge, one is related to the 
universal classes $h$ and the other is related 
to the particles which belong to these classes. For the statistical 
parameter in the interval $0$$\;$$ < $$\;$$\nu$$\;$$ <$$\;$$ 1$ (the first 
elements of each class $h$), $c[\nu]
=i_{f}[h,\nu]$, as otherwise we obtain different values.

In another way, the central charge $c[\nu]$ can be obtained using the 
Rogers dilogarithm function\cite{R6}, i.e.

\begin{equation}
\label{e.16}
c[\nu]=\frac{L[x^{\nu}]}{L[1]},
\end{equation}

\noindent with $x^{\nu}=1-x$,$\;$ $\nu=0,1,2,3,etc.$ and 

\begin{equation}
L[x]=-\frac{1}{2}\int_{0}^{x}\left\{\frac{\ln(1-y)}{y}
+\frac{\ln y}{1-y}\right\}dy,\; 0 < x < 1.
\end{equation}

\noindent Thus, taking into account the Eqs.(\ref{e.11},\ref{e.12}), 
we can extract the sequence of fractal indexes ( Tables II and III ).

\newpage

\vspace{10mm}
\begin{center}
Table II
\end{center}
\begin{center}
\begin{tabular}{|c|c|c|c||||c|c|c|c|}
\hline
$h$ & $\nu$ & $s$ & $i_{f}[h]=c[\nu]$ & $h$ & $\nu$ 
& $s$ & $c[\nu]$\\
\hline
$2$ & $0$ & $0$ & $1$ & $2$ & $0$ & $0$ & $1$ \\
\hline
$\frac{39}{20}$ & $\frac{1}{20}$ & $\frac{1}{40}$ 
& $0.858$ & $2$ & $20$ & $10$ & $0.142$\\
\hline
$\frac{37}{19}$ & $\frac{1}{19}$ & $\frac{1}{38}$ 
& $0.854$ & $1$ & $19$ & $\frac{19}{2}$ & $0.146$\\
\hline
$\frac{35}{18}$ & $\frac{1}{18}$ & $\frac{1}{36}$ 
& $0.849$ & $2$ & $18$ & $9$ & $0.151$\\
\hline
$\frac{33}{17}$ & $\frac{1}{17}$ & $\frac{1}{34}$ 
& $0.845$ & $1$ & $17$ & $\frac{17}{2}$ & $0.155$\\
\hline
$\frac{31}{16}$ & $\frac{1}{16}$ & $\frac{1}{32}$ 
& $0.84$ & $2$ & $16$ & $8$ & $0.16$\\
\hline
$\frac{29}{15}$ & $\frac{1}{15}$ & $\frac{1}{30}$ 
& $0.834$ & $1$ & $15$ & $\frac{15}{2}$ & $0.166$\\
\hline
$\frac{27}{14}$ & $\frac{1}{14}$ & $\frac{1}{28}$ 
& $0.829$ & $2$& $14$ & $7$ & $0.171$\\
\hline
$\frac{25}{13}$ & $\frac{1}{13}$ & $\frac{1}{26}$ 
& $0.822$ & $1$& $13$ & $\frac{13}{2}$ & $0.178$\\
\hline
$\frac{23}{12}$ & $\frac{1}{12}$ & $\frac{1}{24}$ 
& $0.814$ & $2$ &  $12$ & $6$ & $0.186$\\
\hline
$\frac{21}{11}$ & $\frac{1}{11}$ & $\frac{1}{22}$ 
& $0.806$ & $1$ & $11$ & $\frac{11}{2}$ & $0.194$\\
\hline
\end{tabular}
\end{center}
\vspace{10mm}

\vspace{10mm}
\begin{center}
Table III
\end{center}
\begin{center}
\begin{tabular}{|c|c|c|c|||||c|c|c|c|}
\hline
$h$ & $\nu$ & $s$ & $i_{f}[h]=c[\nu]$ & $h$ & $\nu$ 
& $s$ & $c[\nu]$\\
\hline
$\frac{19}{10}$ & $\frac{1}{10}$ & $\frac{1}{20}$ 
& $0.797$ & $2$ & $10$ & $5$ & $0.203$\\
\hline
$\frac{17}{9}$ & $\frac{1}{9}$ & $\frac{1}{18}$ 
& $0.786$ & $1$ & $9$ & $\frac{9}{2}$ & $0.214$\\
\hline
$\frac{15}{8}$ & $\frac{1}{8}$ & $\frac{1}{16}$ 
& $0.774$ & $2$ & $8$ & $4$ & $0.226$\\
\hline
$\frac{13}{7}$ & $\frac{1}{7}$ & $\frac{1}{14}$ 
& $0.759$ & $1$ & $7$ & $\frac{7}{2}$ & $0.241$\\
\hline
$\frac{11}{6}$ & $\frac{1}{6}$ & $\frac{1}{12}$ 
& $0.742$ & $2$ & $6$ & $3$ & $0.258$\\
\hline
$\frac{9}{5}$ & $\frac{1}{5}$ & $\frac{1}{10}$ 
& $0.721$ & $1$ & $5$ & $\frac{5}{2}$ & $0.279$\\
\hline
$\frac{7}{4}$ & $\frac{1}{4}$ & $\frac{1}{8}$ 
& $0.693$ & $2$& $4$ & $2$ & $0.307$\\
\hline
$\frac{5}{3}$ & $\frac{1}{3}$ & $\frac{1}{6}$ 
& $0.656$ & $1$ & $3$ & $\frac{3}{2}$ & $0.344$\\
\hline
$\frac{3}{2}$ & $\frac{1}{2}$ & $\frac{1}{4}$ 
& $0.6$  & $2$ & $2$ & $1$ & $0.4$\\
\hline
$1$ & $1$ & $\frac{1}{2}$ & $0.5$ & $1$ & $1$ 
& $\frac{1}{2}$ & $0.5$\\
\hline
\end{tabular}
\end{center}
\vspace{5mm}

\newpage

\noindent On the one way, we can estimate the fractal index for 
the dual classes of $h$ with rational values, 
considering a fitting of the graphics $i_{f}[h]\times h$ and 
$c[\nu]\times \nu$, plus the observation that 
the $i_{f}[h,\nu]$ diminishes in the sequence

\begin{eqnarray}
i_{f}[h,\nu]&=&i_{f}\left[\frac{3}{2},\frac{1}{2}\right],\;\;\;\;\;\;
i_{f}\left[\frac{4}{3},\frac{2}{3}\right],\;\;\;\;\;
i_{f}\left[\frac{5}{4},\frac{3}{4}\right],\nonumber\\
&&i_{f}\left[\frac{6}{5},\frac{4}{5}\right],\;\;\;\;\;\;
i_{f}\left[\frac{7}{6},\frac{5}{6}\right],\;\;\;\;\;
i_{f}\left[\frac{8}{7},\frac{6}{7}\right],\nonumber\\
&&i_{f}\left[\frac{9}{8},\frac{7}{8}\right],\;\;\;\;\;\;
i_{f}\left[\frac{10}{9},\frac{8}{9}\right],\;\;\;
i_{f}\left[\frac{11}{10},\frac{9}{10}\right],\nonumber\\
&&i_{f}\left[\frac{12}{11},\frac{10}{11}\right],\;\;
i_{f}\left[\frac{13}{12},\frac{11}{12}\right],\;\;
i_{f}\left[\frac{14}{13},
\frac{12}{13}\right],\nonumber\\
&&i_{f}\left[\frac{15}{14},\frac{13}{14}\right],\;\;
i_{f}\left[\frac{16}{15},
\frac{14}{15}\right],\;\;
i_{f}\left[\frac{17}{16},\frac{15}{16}\right],\nonumber\\
&&i_{f}\left[\frac{19}{18},\frac{17}{18}\right],\;\;
i_{f}\left[\frac{20}{19},\frac{18}{19}\right],\;\;
i_{f}\left[\frac{21}{20},
\frac{19}{20}\right],\nonumber\\
&&i_{f}\left[1,1\right].\nonumber
\end{eqnarray}


This way, we observe that our formulation to 
the universal class $h$ of particles with any values of spin $s$ 
establishes a connection between Hausdorff dimension $h$ and 
the central charge $c[\nu]$, in a manner unsuspected till now. 
Besides this, we have obtained a connection between $h$ and the 
Rogers dilogarithm function, through the fractal index defined 
in terms of the partition function associated with the universal 
class $h$ of particles. Thus, considering 
the Eqs.(\ref{e.11}, \ref{e.12}) and the Eq.(\ref{e.16}), we have

\begin{eqnarray}
\frac{L[x^{\nu}]}{L[1]}&=&
i_{f}[h,\nu]-i_{f}\left[h,\frac{1}{\nu}\right],\; 
\nu=0,2,4,etc.\\
\frac{L[x^{\nu}]}{L[1]}&=&
2\times i_{f}[h,\nu]-i_{f}\left[h,\frac{1}{\nu}\right],\;
\nu=1,3,5,etc.
\end{eqnarray}

 Also in\cite{R1} we have established 
a connection between the fractal parameter $h$ and the Farey 
series of rational numbers, therefore once the classes $h$ satisfy all the 
properties of these series we have an infinity 
collection of them. In this sense, we clearly establish a connection 
between number theory and Rogers dilogarithm function. Given that 
the fractal parameter is an irreducible number $h=\frac{p}{q}$, 
the classes satisfy the properties\cite{R7}

P1. If $h_{1}=\frac{p_{1}}{q_{1}}$ and 
$h_{2}=\frac{p_{2}}{q_{2}}$ are two consecutive fractions 
$\frac{p_{1}}{q_{1}}$$ >$$ \frac{p_{2}}{q_{2}}$, then 
$|p_{2}q_{1}-q_{2}p_{1}|=1$.

P2. If $\frac{p_{1}}{q_{1}}$, $\frac{p_{2}}{q_{2}}$,
$\frac{p_{3}}{q_{3}}$ are three consecutive fractions 
$\frac{p_{1}}{q_{1}}$$ >$$ \frac{p_{2}}{q_{2}} 
$$>$$ \frac{p_{3}}{q_{3}}$, then 
$\frac{p_{2}}{q_{2}}=\frac{p_{1}+p_{3}}{q_{1}+q_{3}}$.

P3. If $\frac{p_{1}}{q_{1}}$ and $\frac{p_{2}}{q_{2}}$ are 
consecutive fractions in the same sequence, then among 
all fractions\\
 between the two, 
$\frac{p_{1}+p_{2}}{q_{1}+q_{2}}$
 is the unique reduced
fraction with the smallest denominator.

For example, consider the Farey series of order 6, denoted by 
the $\nu$ sequence

\begin{eqnarray}
(h,\nu)&=&\left(\frac{11}{6},\frac{1}{6}\right)\rightarrow 
\left(\frac{9}{5},\frac{1}{5}
\right)\rightarrow
\left(\frac{7}{4},\frac{1}{4}\right)\rightarrow
 \left(\frac{5}{3},\frac{1}{3}\right)\rightarrow\nonumber\\
&&\left(\frac{8}{5},\frac{2}{5}\right)\rightarrow 
\left(\frac{3}{2},\frac{1}{2}\right)
\rightarrow \left(\frac{7}{5},\frac{3}{5}\right)
\rightarrow 
\left(\frac{4}{3},\frac{2}{3}\right) \rightarrow\\ 
&&\left(\frac{5}{4},\frac{3}{4}\right) \rightarrow 
\left(\frac{6}{5},\frac{4}{5}\right) \rightarrow 
\left(\frac{7}{6},\frac{5}{6}\right) \rightarrow 
\cdots.\nonumber
\end{eqnarray}

\noindent Using the fractal spectrum ( Eq.\ref{e.7} ), we can obtain other 
sequences which satisfy the Farey properties and for the classes
 
\[
h=\frac{11}{6},\frac{9}{5},
\frac{7}{4},\frac{5}{3},
\frac{8}{5},\frac{3}{2},
\frac{7}{5},\frac{4}{3},
\frac{5}{4},\frac{6}{5},
\frac{7}{6},\cdots,
\] 

and ( note that these ones are dual classes, ${\tilde{h}}=3-h$ ) 
we can calculate the fractal 
index taking into account the Rogers dilogarithm function or the 
partition function associated with each $h$.

Now, in\cite{R1} we also considered free fractons and 
an equation of state at low temperatures was obtained

\begin{equation}
P=\frac{h\rho^2}{2\gamma}+\gamma(KT)^2{\cal{C}}_{1}(h),
\end{equation}

\noindent where $\gamma=\frac{m(\nu+1)}{4\pi\hbar^2}$ ( $\hbar$ 
is the Planck constant ), $\rho$ is the particle density and 

\begin{equation}
{\cal{C}}_{1}(h)=-\int_{1(T=0)}^{\infty(T=\infty)}\frac{d{\cal{Y}}}
{({\cal{Y}}-1)({\cal{Y}}-2)}
\ln\left\{\frac{{\cal{Y}}-1}{{\cal{Y}}-2}\right\}=\frac{\pi^2}{6}.
\end{equation}

\noindent Thus, for the fracton systems we obtain the specific heat $C$ as 

\begin{equation}
\frac{C}{L^2}=\frac{m}{4\pi\hbar^2}K^2 T(\nu+1)\frac{\pi^2}{3}.
\end{equation}

On the other hand, the specific heat of a conformal 
field theory is given by\cite{R5}

\begin{equation}
\frac{C}{L}=\frac{1}{2\pi\hbar v}K^2 T\frac{\pi^2}{3}c[\nu].
\end{equation}

\noindent Comparing the expressions, we obtain the Fermi velocity as

\begin{equation}
v\sim\frac{c[\nu]}{\nu+1},
\end{equation}

\noindent so for $\nu=0$$,$$\;$ $c[0]=1$$,$$\;$ $v\sim 1$$;
$$\;$ $\nu=1$$,$
$\;$ $c[1]=
\frac{1}{2}$$,$$\;$ 
$v\sim 0.25 $$;$$\;$$\;$ $\nu=2$$,$
$\;$ $c[2]=
\frac{2}{5}$$,$$\;$ 
$v\sim 0.133$$;$$\;$ $\nu=3$$,$$\;$ $c[3]=0.344$$,$$\;$ $v\sim 0.086$$;$
$\;$ etc. We observe that fractons are objects defined in 2+1-dimensions 
( see \cite{R1} for more details ).

In summary, we have obtained a connection 
between fractons and CFT-quasiparticles. This was implemented with 
the notion of the {\it fractal index} associated with 
the universal class $h$ of the fractons. This way, 
fractons and CFT-quasiparticles satisfy a specific 
fractal statistics. We also have obtained an expression 
for the Fermi velocity in terms of the conformal anomaly 
and the statistical parameter\cite{R8}. A connection between the Rogers 
dilogarithm function, Farey series of rational numbers and 
Hausdorff dimension $h$ also was established. The idea of fractons as 
quasiparticles has been 
explored in the contexts of the fractional quantum Hall effect\cite{R1}, 
Luttinger liquids\cite{R8} and high-$T_{c}$ superconductivity\cite{R9}. Finally, 
a connection between fractal statistics and 
black hole entropy also was exploited in\cite{R10} and a 
fractal-deformed Heisenberg 
algebra for each class of fractons was introduced in\cite{R11}.

\acknowledgments

We would like to thank an anonymous referee by the comments.

\end{document}